\documentclass{article}
\usepackage{spconf,amsmath,graphicx}

\title{Neural Network Segmentation of Cell Ultrastructure Using Incomplete Annotation}
\name{John Paul Francis$^{1,2}$, Hongzhi Wang$^3$, Kate White$^2$, Tanveer Syeda-Mahmood$^3$, Raymond Stevens$^2$}
\address{$^1$Department of Computer Science, University of Southern California, Los Angeles, California\\
$^2$Department of Biological Sciences, University of Southern California, Los Angeles, California\\
$^3$IBM Almaden Research Center, San Jose, California}
\begin{document}
%
\maketitle
\begin{abstract}
The Pancreatic beta cell is an important target in diabetes research. For scalable modeling of beta cell ultastructure, we investigate automatic segmentation of whole cell imaging data acquired through soft X-ray tomography. During the course of the study, both complete and partial ultrastrucutre annotations were produced manually for different subsets of the data. To more effectively use existing annotations, we propose a method that enables the application of partially labeled data for full label segmentation. For experimental validation, we apply our method to train a convolutional neural network with a set of 12 fully annotated data and 12 partially annotated data and show promising improvement over standard training that uses fully annotated data alone.
\end{abstract}
\begin{keywords}
Pancreatic beta cell, soft x-ray tomography, convolutional neural networks, semantic segmentation
\end{keywords}

\begin{figure*}
\centering
\noindent\includegraphics[height=3.5cm]{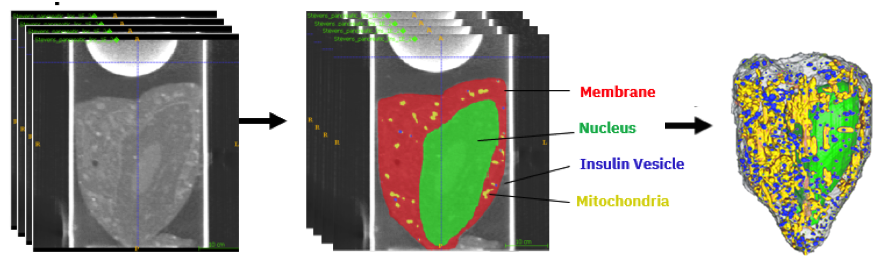}
\caption{The original tomogram (left) is a 512x512x512, 8-bit grayscale image, acquired at a resolution of $\sim$25nm$^3$/voxel. During segmentation, this image is labeled with a 512x512x512 mask containing discrete values [0, 1, 2, (3), and (4)], where 0 corresponds to background, 1 to membrane, 2 to nucleus, 3 to insulin vesicle, and 4 to mitochondria. Values 3 and 4 only exist in labels of INS-1E cells. With these masks, we can then reconstruct a 3D surface model of the cell for visualization and analysis.} 
\label{fig:example}
\end{figure*}

\section{Introduction}
\label{sec:intro}

``Structure determines function" is a central tenet of biology\cite{anfinsenStructFn}. Understanding the structure and function of cells is increasingly important as many of the most pressing modern diseases such as cancer, Alzheimer's, and diabetes are rooted in malfunction at the cellular scale. 

The pancreatic beta cell is responsible for secreting insulin and is therefore an important target in diabetes treatment. To better understand the process of insulin secretion, developing methods to analyze and classify the three-dimensional spatial organization of insulin vesicles along the course of the disease has been an active research topic \cite{granule1, granule2, granule3}. Of additional interest are the relative organization of cellular structures such as the mitochondria, membrane, and nucleus. These features together constitute a subset of the cell's ultrastructure. Observing ultrastructural rearrangements and cell-to-cell variability induced by incretins and other molecules used to treat diabetes can provide mechanistic insights into therapeutic and off-target actions as more effective treatments are being developed.

To model ultastructure of the beta cell, the soft X-ray tomography technique has been successfully applied to acquire 3D cellular imaging data \cite{ekmanMesoscale}. The technique allows for the reconstruction of 3D grayscale images at  approximately 25nm$^3$/voxel (see Fig. \ref{fig:example}) which is sufficient to reveal ultrastructural components in vitro. In order to analyze pathological variance in ultrastructure these individual components must be identified and segmented quantitatively. This process is not trivial and requires significant manual overhead; manually segmenting just one tomogram for membrane, nucleus, mitochondria, and insulin vesicles, for instance, takes experts up to 8 hours and requires specialized software. Due to its high cost, manual annotation is prohibitive for large scale study. Automating this segmentation process is therefore of high interest.

Convolutional neural networks (CNN) have been successfully applied for segmentation in various biomedical imaging domains \cite{ronneberger2015u,litjens2017survey}. However, the excellent segmentation performance achieved by CNNs often relies on the availability of a large amount of fully annotated training data.  In this work, we investigate the application of CNN segmentation for the pancreatic beta cell including incompletely annotated training data to increase our training set and improve overall annotation performance. 

Over the course of our project, tomograms of whole beta cells were collected from three different cell lines: the 1.1B4 human beta cell line, the human embryonic kidney (HEK) cell line and the INS-1E cell line. 1.1B4 cells and HEK cells were collected in the early stage of our project. For these cells, only the membrane and nucleus were annotated and insulin vesicles and mitochondria were labeled as part of the membrane. INS-1E cells were collected more recently. For these cells, more detailed annotations of insulin vesicles and mitochondria inside the membrane were produced in addition to the membrane and nucleus. Since the ``membrane" label has different semantic meaning in the partially labeled data versus the fully annotated data, mixing the partially annotated data with fully annotated data in a classical learning scheme will undermine the performance for those labels with inconsistent meanings. Rather than discard incompletely labeled data, we propose to address this problem with a method that can learn effectively with such incomplete annotations. 

Our method is based on the observation that each partially annotated data can be derived from fully annotated data by merging labels. In our application, partially labeled data for 1.1B4 and HEK cells can be derived from fully annotated data by merging insulin vesicle and mitochondria labels with the membrane label. Following this derivation, any fully annotated data output by our network can be compared with partially labeled manual annotation in training. 

In a more general context, it is common to have such incomplete annotations produced over projects with different objectives, and our method can be applied for any incompletely labeled data so long as such a merging procedure from full labels to partial labels is possible. For instance, a study focusing on lung diseases may have produced annotations for only lung structures using chest CT volumes, while a study on cardiac vascular diseases may have produced annotations for only cardiac structures using a different set of chest CT volumes. In this example, both the lung-only annotation and the cardiac-only annotation are partially labeled data that could be derived from a full annotation containing both lung and cardiac structures by merging the other label with the background. After merging, a fully annotated output can always be trained against partial ground truth in training. Furthermore, since each anatomical structure may provide contextual information for other anatomical structures, learning with various partially labeled data not only informs a unified segmentation model for each anatomical structure alone, but also improves the overall segmentation performance for every structure together. 

Note that learning with incomplete annotation is different from learning with sparse annotation \cite{cciccek20163d}, where only part of the training data is fully annotated and the rest is completely unlabeled, and learning with ambiguous annotation \cite{cour2011learning}, where a set of candidate labels are given for each instance but only one label is correct. Learning with incomplete annotation addresses unique situations in which annotations contain only a subset of labels of interest which subsume missing, more detailed labels of greater semantic depth.

In experimental validation, combining 12 partially annotated data with 12 fully annotated data for CNN training produced substantial improvement for every anatomical structure over using the fully annotated data alone.

\section{Methods}
\label{sec:method}

\subsection{Data Collection and Processing}
Soft x-ray tomography was performed on 216 INS-1E cells. 27 of those resulting tomograms were manually segmented for membrane, nucleus, mitochondria, and insulin vesicles (See fig. \ref{fig:example}). In addition, soft x-ray tomography was performed on HEK and 1.1B4 cells, 12 of which were segmented by experts as part of a negative control group. For HEK and 1.1B4 cells, insulin vesicles and mitochondria were labeled as part of membrane. It should be noted that while this inconsistency in our manually segmented data was an artifact of our particular project, diverse data sets from similar imaging conditions are common in general and the ability to leverage heterogeneous data in neural network training presents many opportunities, especially when labeled data is scarce (See Discussion in Section 4).

The original image and manual segmentation for each annotated tomogram were resized to a standard 512x512x512 voxels using linear and nearest-neighbor interpolation, respectively. From the 27 fully annotated INS-1E data, 12 were randomly selected for training. 5 were selected for validation and 10 were selected for testing. For our combined method, all 12 partially annotated 1.1B4/HEK cells were included for model training as well.

\subsection{Learning with Partially Labeled Data}
Let $\mathcal{L}$ be a complete label set. Without loss of generality, we assume that each voxel in an image is assigned with a single label. Let $\mathcal{I}^F=\{(I_i,S_i)\}_{i=1}^n$ be $n$ fully labeled training images, where $I_i$ and $S_i$ are image and ground truth segmentation, respectively. Let $\mathcal{I}^P=\{(I_j,S_j^P)\}_{j=1}^m$ be $m$ partially labeled training images, with label set $\mathcal{L}^P$ and $|\mathcal{L}^P|<|\mathcal{L}|$. Let $T^P:\mathcal{L}\rightarrow \mathcal{L}^P$ be a mapping (label merging) function that maps each label from the full label set $\mathcal{L}$ to one label in $\mathcal{L}^P$. Note that the label in $\mathcal{L}^P$ may have different semantic meaning, i.e. union of multiple labels, from the same label in $\mathcal{L}$. In our application, $T^P$ is an identity function except $T^P(\mbox{mitochondria})=\mbox{membrane}$ and $T^P(\mbox{insulin vesicle})=\mbox{membrane}$.

Let $\theta$ be the parameters of a convolutional neural network. Let $M_{\theta}(I)$ be the result produced by the model for image I. We propose the following objective function for combining fully and partially labeled data in training:
\begin{equation}\label{e:formulation}
L(\theta) = \sum_{i=1}^n L^{F}(M_{\theta}(I_i),S_i)+\sum_{j=i}^mL^{P}(M_{\theta}(I_j), S_j^P)
\end{equation}
$L^{F}$ is the standard term on fully labeled training data. $L^{P}$ is applied on partially labeled data. For training data with full annotation, we apply the cross entropy function \cite{lin2017focal}.
\begin{equation}
L^F(M_{\theta}(I),S)=\sum_{l\in\mathcal{L}}\sum_{x\in r(S,l)} -  \log(p_l(x|I,\theta))\label{e:crossentropy}
\end{equation}
$x$ index through image voxels. $p_l$ is the model predicted probability for label $l$ from the full label set $\mathcal{L}$. $r(S,l)$ is the region assigned to $l$ in ground truth $S$. 

For partially label training data, we apply a modified cross entropy function by transforming model predications for label set $\mathcal{L}$ to model predictions for label set $\mathcal{L}^P$:
\begin{equation}
L^P(M_{\theta}(I),S^P)=\sum_{l\in\mathcal{L}^P}\sum_{x\in r(S^P,l)} -  \log(p^P_l(x|I,\theta))\label{e:crossentropy_parital}
\end{equation}
where $p^P$ is the model predicted probability for label set $\mathcal{L}^P$, which is derived from the model predicted probability for $\mathcal{L}$ by:
\begin{equation}
p^P_l(x|I,\theta) = \sum_{T^P(l')=l}p_{l'}(x|I,\theta)
\end{equation}

In our application, only one partially labeled dataset is available. In a more general scenario, multiple partially labeled datasets may be available from different data sources or from different research projects. For those scenarios, each partially labeled dataset can be represented by a distinct label mapping function $T^P$. Our formulation can be applied to include each partially labeled dataset for training by adding its corresponding $L^P$ term to the objective function.

\subsection{Network Architecture}
Due to the memory constraint, we chose a standard 2D U-net architecture \cite{ronneberger2015u} instead of a 3D CNN. Our choice is based on the consideration that both mitochondria and insulin vesicles are small scale structures. Downsampling 3D volumes to fit GPU memory may therefore discard necessary detail and compromise segmentation performance for small structures. Alternatively, a patching scheme separating the volume into smaller sub-blocks to fit GPU memory would deprive the model of larger scale context for prediction. Our 2D network consisted of a contracting path and an expanding path. The contracting path was 5 levels deep with 2$\times$2 pooling between each level, while the expanding path was connected by an upsampling filter with a 2$\times$2 kernel. All convolutions had a kernel size of 3$\times$3, stride=1, pad=1 followed by rectified linear units (ReLu). Each level was composed of 2 convolutions back to back. The number of filters at the top level was 32 and doubled at each level of depth in the network. The last layer contained 1$\times$1 convolution followed by a softmax, which gave the pixel-wise probabilities of each segmentation label.

\begin{table*}
\centering
\begin{tabular}{ |p{2cm}p{2cm}||p{2cm}|p{2cm}|p{2cm}|p{2cm}|p{2cm}|}
 \hline
 \multicolumn{7}{|c|}{DSC} \\
 \hline
 &  &   Membrane    &   Nucleus &   Insulin    &Mitochondria   &   Average\\
 \hline
 Validation Set &   Baseline   & 0.848 &   0.928   &   0.515   &   0.458   &   0.687\\
                &   Mixed Labels & 0.889&   0.917   &   0.565   &   0.584   &   0.739\\

 &&&&&&\\
 Testing Set    &   Baseline  & 0.813  &   0.718   &   0.559   &   0.467   &   0.639\\
                & Mixed Labels & 0.843 &   0.817    &   0.588   &   0.575   &   0.706\\
        & \textit{Difference in percentage} & \textit{+3.6\%}   &   \textit{+13.8\%}   &   \textit{+5.2\%}   &  \textit{+23.1\%} &   \textit{\textbf{+10.5\%}}  \\
 \hline
\end{tabular}
\caption{Quantitative dice scores evaluated on the validation and testing sets. The mixed label method outperformed the baseline method at every level in every category except for the Nucleus label in the validation. In that specific case, both methods performed very well and the relative decline of ~1\% was likely an artifact of the smaller sized validation set.} \label{tab:sometab}
\end{table*}

\begin{figure}
        \centering
            \includegraphics[width=8.5cm]{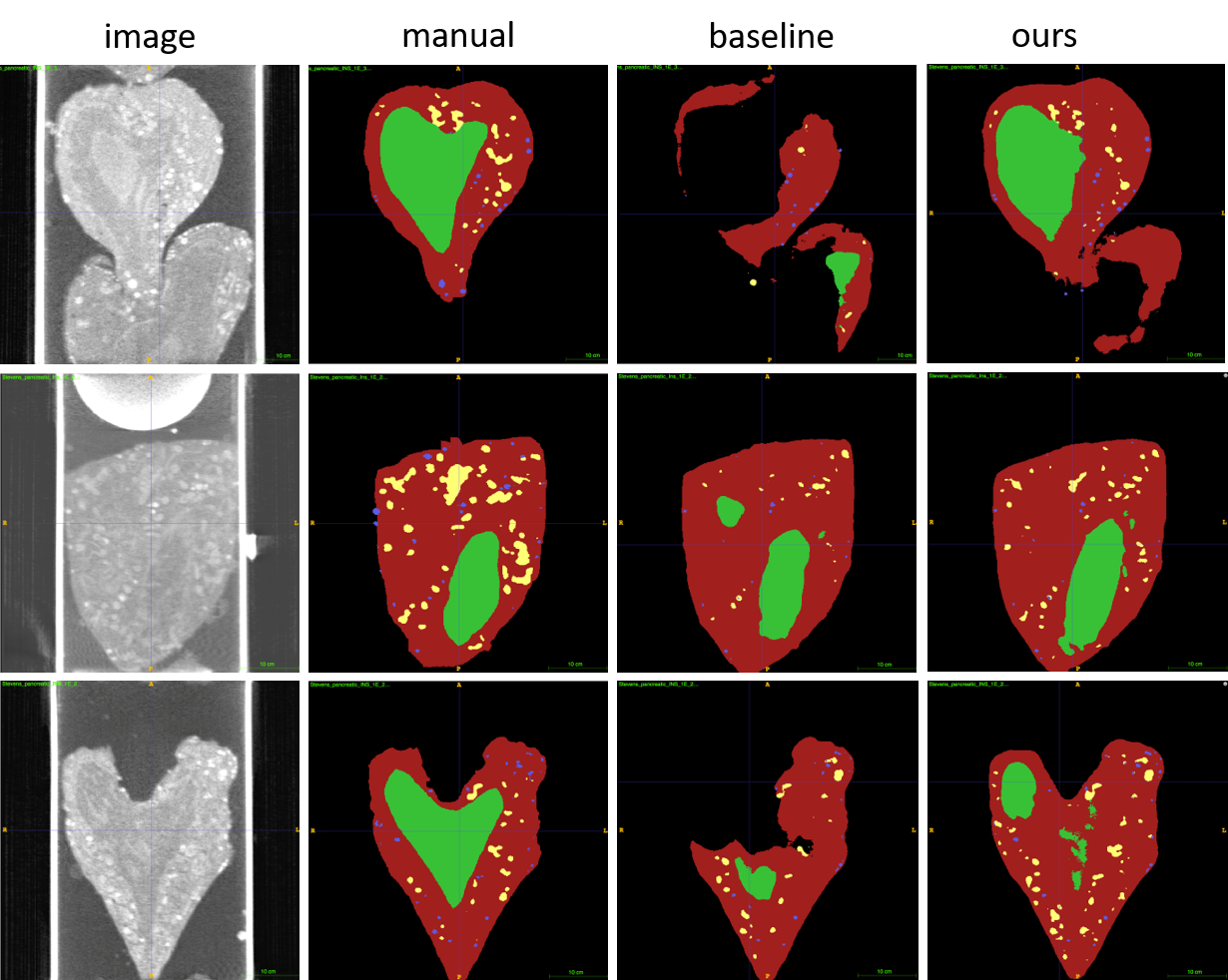}
        \caption{Example segmentation results on the testing set. Including partially annotated data for training demonstrates better alignment with the ground truth, more thorough labeling on inner structures, and more robustness to varying backgrounds in the image.}
        \label{fig:results}
\end{figure}

\section{Experiments}
\label{sec:experiment}
For our method, both the fully annotated and partially labeled training data were applied for CNN training. For comparison, the baseline method was trained only using the fully annotated training data. In both cases, data augmentation was implemented by applying a randomly generated affine transform, with (-180$^o$, 180$^o$] rotation and [-20\%, 20\%] scale variation, to the training data on-the-fly. Both methods were run for 200 epochs. For quantitative evaluation, the dice similarity coefficient (DSC) \cite{Dice45} was applied to measure agreement between automatic and manual segmentation.

Fig. \ref{fig:results} shows example segmentation results produced for testing data for each segmentation method. Table 1 summarizes the segmentation performance on validation and testing data, respectively. As expected, including partially annotated training data for training substantially improved segmentation performance for membrane and nucleus, which are labeled in both fully and partially annotated data. Interestingly, prominent improvement was also observed for insulin vesicles and mitochondria, which are not labeled in the partially labeled training set. One potential explanation for this result is that including partially labeled data for training doubled labeled training data, which is critical for learning a more robust overall representation for better segmentation. The improvement for these labels also suggests that a more robust knowledge of the distribution on one set of structures in the cell can inform the inference of others. Specifically, both insulin vesicle and mitochondria are located within membrane and if a segmentation method misclassifies membrane regions as other structures, it is likely to misclassify insulin vesicle and mitochondria as well. 

Overall, the segmentation performance was improved from 0.639 DSC to 0.706 DSC, a 10.5\% improvement over the baseline performance. Our results clearly demonstrate the advantage of combining data with incomplete annotations in CNN training. 

\section{Discussion and Conclusion}
\label{sec:typestyle}

In this paper we outlined the motivation for modeling the structure of the pancreatic beta cell as well as the need to segment ultrastructural features towards that end. We presented the need to automate this process of segmentation as well as the challenges cellular researchers face in achieving it, namely due to a lack of high numbers of completely labeled data. By allowing for inconsistency in annotations, we show that diverse annotation sets can be combined in training to yield significant improvement in performance and clearly demonstrate the tenability of combining diverse annotation sets for segmentation in general.

It is sensible that areas which require manual work most are often the least prepared to receive automation out of the box. In machine learning research, techniques are often developed on well formatted data sets in which labels are easy to acquire and the number of labeled samples is not a limiting variable. In practice, however, and especially in biomedical research, annotation is not trivial and labeled data is limited. More widespread techniques to combine existing heterogenous annotations have the potential to make use of sparse and increasingly diverse datasets, thereby making neural networks more tractable for specialized applications on the path to curing diabetes and beyond.

It should also be mentioned that although part of the training data was fully annotated in our application, fully annotated data is not necessarily required. Indeed, even our "fully" annotated data is only partial, and acquiring truly complete annotated data of a cell is likely beyond the reasonable capacity of any person alone. Learning a unified segmentation model from multiple partially annotated data will be an interesting application moving forward.

\section{Acknowledgdments}
Soft x-ray tomography data were collected at The National Center for X-ray Tomography, which is supported by the National Institute of General Medical Sciences of the National Institutes of Health Grant P41GM103445 and the US Department of Energy, Office of Biological and Environmental Research Grant DE-AC02-05CH11231

\bibliographystyle{IEEEbib}
\bibliography{strings,refs}

\end{document}